\documentclass[pra,twocolumn,showpacs,superscriptaddress,nofootinbib]{revtex4-1}
\usepackage{amssymb}
\usepackage{amsbsy}
\usepackage{amsmath}
\usepackage[pdftex]{graphics}
\usepackage{graphics}
\usepackage{graphicx}

\usepackage[colorlinks,linkcolor=blue]{hyperref}

\usepackage{mathptmx} 

\def\q{{\rm\bf q}}
\def\k{{\rm\bf k}}

\def\la{\langle}
\def\ra{\rangle}

\def\Om{\Omega}

\def\tr{\rm{Tr}}

\newcommand{\beq}{\begin{equation}}
\newcommand{\eeq}{\end{equation}}
\newcommand{\beqa}{\begin{eqnarray}}
\newcommand{\eeqa}{\end{eqnarray}}

\begin{document}
\title{Shortcuts to adiabaticity in a time-dependent  box}
\author{A. del Campo$^{1,*}$, M. G. Boshier$^2$}
\affiliation{Los Alamos National Laboratory, Theoretical Division  T-4 and T-CNLS,  Los Alamos, NM 87545, USA}
\affiliation{Los Alamos National Laboratory, Physics Division, Los Alamos, NM 87545, USA}

\begin{abstract}

A method is proposed to drive an ultrafast non-adiabatic dynamics of an ultracold gas trapped in a time-dependent box potential. The resulting state is free from spurious excitations associated with the breakdown of adiabaticity, and preserves the quantum correlations of the initial state up to a scaling factor.
The process relies on the existence of an adiabatic invariant and the inversion of the dynamical self-similar scaling law dictated by it. Its physical implementation generally requires the use of an auxiliary expulsive potential. The method is extended to a broad family of interacting many-body systems. As illustrative examples we consider the ultrafast expansion of a Tonks-Girardeau gas and of Bose-Einstein condensates in different dimensions, where the method exhibits an excellent robustness against different regimes of interactions and the features of an experimentally  realizable  box potential.

\end{abstract}


\maketitle
\noindent


``Fast good'' is a new culinary concept envisioned by the chef F. Adri\`a aimed at creating a diet enjoying of two apparently mutually-exclusive features: fast-service and high-quality.
When similar ideas are invoked in the quantum realm, one faces the adiabatic theorem which imposes a price. The preparation of a given target state with high-fidelity, free from spurious excitations, generally demands
a long time of evolution and the implementation of an adiabatic dynamics.
As a result, it comes as no surprise that the recent development of shortcuts to adiabaticity (STA) \cite{chen10}, has led to a surge of theoretical  \cite{chen10b,Muga,delcampo10,delcampo11,transport1,transport2,control1,control2,COS11} and experimental activity \cite{Labeyrie10,Labeyrie10b,Morsch}.

It is the purpose of this work to show that STA can be implemented in a time-dependent box, the paradigmatic model of a quantum piston \cite{QJ12}.
The relevance of this confinement is enhanced by the development of experimental techniques to create optical billiards for ultracold gases \cite{prepainters1,prepainters2}, 
 paint arbitrary potential shapes for Bose-Einstein condensates \cite{painters},  and realize  all-optical boxes
 \cite{becbox} and analogous traps in atom chips \cite{boxchip}. It has the potential to greatly advance the field of quantum simulation, 
facilitating the connection between ultracold atom experiments and condensed matter systems. 

The prospects of finding STA for  box confinement seem challenging at the very least, based on the following considerations. 
Since the early insight by  M. Moshinsky in 1952, it is known that the rapid expansion  of matter-waves initially localized in a region of space exhibits quantum transients and excitations in the form of density ripples, a phenomenon referred to as diffraction in time \cite{Moshinsky52, GK76,DGCM09}. Moreover, reflections from the walls of the trap generally lead to Talbot oscillations and the formation of quantum carpet in the time evolution of the density profile \cite{Schleich1,Schleich2}. On top of that, a series of work during the last decades have shown that the suppression of excitations in an expanding box is inevitably constrained by the adiabatic theorem both in the non-interacting \cite{boxlaws1,boxlaws2,boxlaws3,bookinv} and mean-field regime \cite{BMT02}. 

In spite of these results preventing ultrafast excitation-free dynamics in box traps,  we show that
fast non-adiabatic expansion or compression is in fact possible,  allowing preparation in a predetermined finite time of the same target state than the adiabatic evolution.
The key to  this shortcut to adiabaticity is the use of an auxiliary potential term superimposed on the box trap.  

We shall start by introducing a dynamical invariant in a time-dependent box trap at the single-particle level, and use it  to derive a scaling law that governs the time evolution. The inversion of this scaling law will be the key to engineer the STA. This result will be extended to  many-body systems with a broad family of interactions.  Finally, we shall illustrate the method with some  examples relevant to experiments with ultracold gases, showing that the STA is robust in the presence of interactions and  experimental imperfections.


 \noindent
{\bf Dynamical invariants and self-similar dynamics} 

For a time-dependent box of width $\xi(t)$,  the scaling laws governing the dynamics of the expanding eigenstates reported to date are associated with trajectories of the form $\xi(t)=[at^2+bt+c]^{\frac{1}{2}}$  (with $a$, $b$, $c$ real constants)  \cite{boxlaws1,boxlaws2,boxlaws3,bookinv} , which turn out to be unsuited for engineering a STA  (see below).
Nonetheless, given a  time-dependent Hamiltonian $\hat{H}(t)$, it is possible to build a dynamical invariant \cite{LR69}, $\hat{\mathcal{I}}(t)$ such that 
\beqa
\label{ice}
\frac{d\hat{\mathcal{I}}(t)}{dt}=\frac{\partial \hat{\mathcal{I}}(t)}{\partial t}+\frac{1}{i\hbar}[\hat{\mathcal{I}}(t),\hat{H}(t)]=0,
\eeqa
with spectral decomposition $\hat{\mathcal{I}}(t)=\sum_n\lambda_n|\phi_n(t)\ra\la\phi_n(t)|$ in terms of the
set of eigenmodes $|\phi_n(t)\ra$ with eigenvalues $\lambda_n$. This is a particularly useful 
basis to describe the time evolution of an initial state $\Psi$, by the superposition $\Psi(x,t)=\sum_n\exp(i\alpha_n)|\phi_n(t)\ra$, 
where the Lewis-Riesenfeld phase is given by
$\alpha_n=\int_0^t\la\phi_n(t')|i\hbar\partial_{t'}-\hat{H}(t')|\phi_n(t')\ra dt'/\hbar$,
and can be understood as the sum of the dynamical phase and the Aharanov-Anandan phase \cite{bookinv}.
For a time-dependent box of width $\xi(t)$ and initial width $\xi(0)=\xi_0$, a dynamical invariant exists \cite{CL99},
\beqa
\hat{\mathcal{I}}=\frac{1}{2m}\frac{\xi^2(t)}{\xi^2_0}\left(p-m\frac{\dot{\xi}(t)}{\xi(t)}x\right)^2
\eeqa
with eigenvectors $\la x|\phi_n(t)\ra=[2/\xi(t)]^{\frac{1}{2}}\exp\Big[i\frac{m\dot{\xi}(t)}{2\hbar\xi(t)}x^2\Big]\sin[n\pi x/\xi(t)]$ and eigenvalues $\lambda_n=\frac{\hbar^2k^2_n}{2m}$, with $k_n=n\pi/\xi_0$.
The Lewis-Riesenfeld phase can be computed to be
$
\alpha_n(t)=-\frac{\hbar n^2\pi^2}{2m\xi_0^2}\eta(t)
$
with $\eta(t)=\int_0^tdt'\xi_0^2/\xi^2(t')$.
The condition in Eq. (\ref{ice}) for $\hat{\mathcal{I}}$ to be an invariant requires the box potential to be supplemented with an auxiliary harmonic term
\beqa
U^{\rm aux}(x,t)=-\frac{1}{2}m\frac{\ddot{\xi}(t)}{\xi(t)}x^2,
\label{Uaux}
\eeqa
whose frequency $\Omega(t)=\sqrt{-\frac{\ddot{\xi}(t)}{\xi(t)}}$ is dictated by the ratio of the acceleration of the trajectory $\xi(t)$, and the trajectory itself.
Note that $\xi(t)>0$, so that for $\ddot{\xi}(t)>0$, $\Omega(t)$ is purely imaginary, and the auxiliary term $U^{\rm aux}$ is a repulsive harmonic potential.
If $\dot{\xi}(0)=0$,  $[\hat{H}(0),\hat{\mathcal{I}}(0)]=0$ so $\hat{H}(0)$ and $\hat{\mathcal{I}}(0)$ have common eigenstates. Further, if $\ddot{\xi}(0)=0$ holds as well,
then $U^{\rm aux}(x,0)=0$ and an eigenstate $\Psi_n(x,0)$ of the box at $t=0$ evolves into $\Psi_n(x,t)=\exp[i\alpha_n(t)]\phi_n(t)$, a key observation to engineer a STA as we shall see.
We note that the experimental implementation of $U^{\rm aux}(x,t)$ can be assisted by the same techniques used to create the box potential:   the use of a blue-detuned laser \cite{Khaykovich} or direct painting of the required trap \cite{prepainters1,prepainters2,painters}.


\noindent
{\bf Shortcuts to adiabaticity: inverting the dynamical scaling law}

We next discuss how to implement a non-adiabatic expansion of the box by a factor
$\gamma(\tau)=\xi(\tau)/\xi_0$ in a given finite-time $\tau$ suppressing excitations in the final state. We shall impose the condition $U^{\rm aux}(x,0)=U^{\rm aux}(x,\tau)=0$. 
As in the adiabatic case, in a STA the time evolution of an eigenmode of the initial box should
reproduce an eigenmode of the final trap. As at $t=0$, this can be enforced by imposing the condition $\dot{\xi}(\tau)=\ddot{\xi}(\tau)=0$.
%
\begin{figure}[t]
\begin{center}
\includegraphics[width=0.85\linewidth]{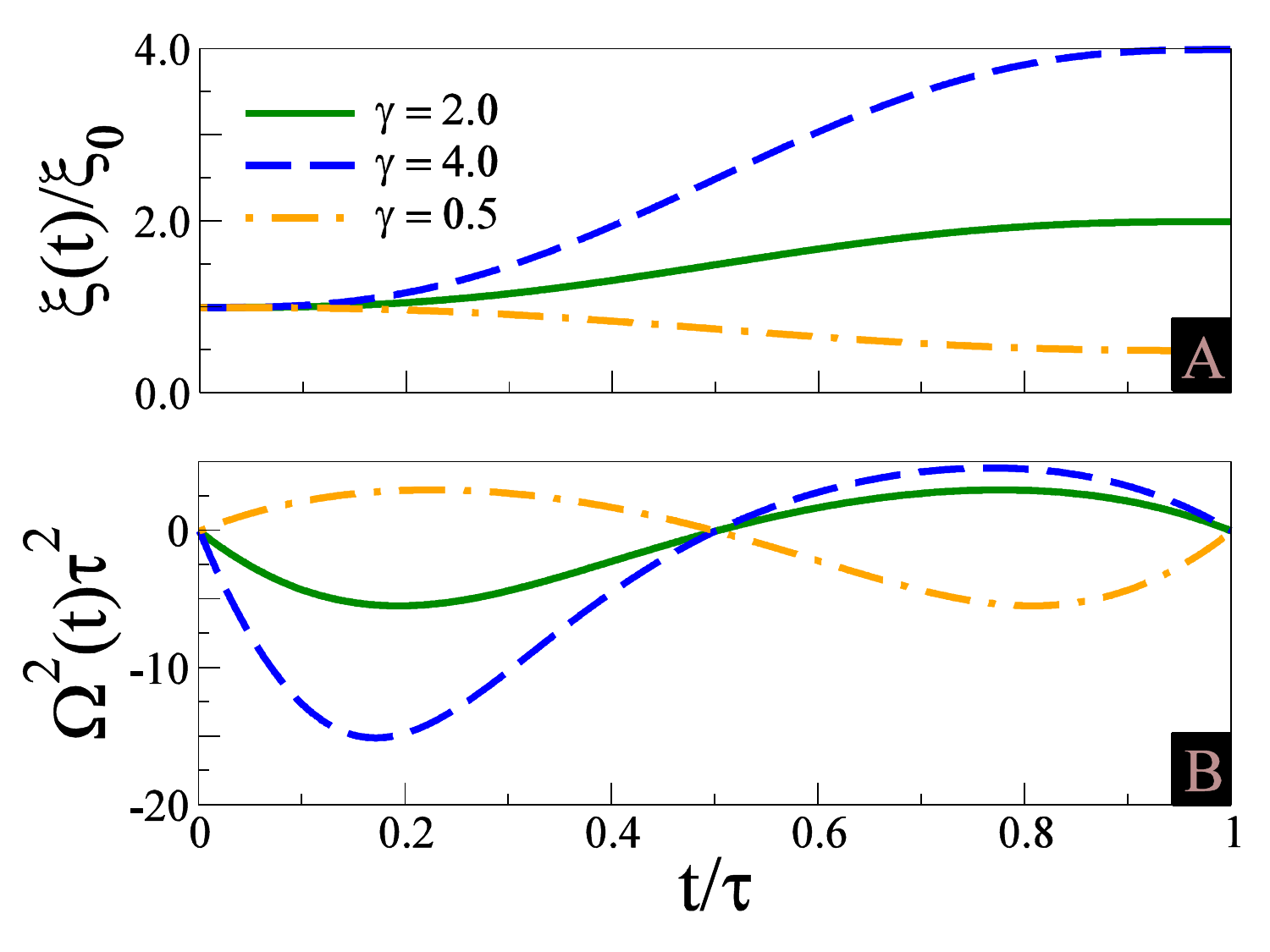}
\end{center}
\caption{Time-modulation of the potential trap along a shortcut to adiabaticity.  (A) Evolution of the box width $\xi(t)$ of a time-dependent box during a shortcut to different adiabatic expansions and a compression in a finite time $\tau$. (B) Frequency of the auxiliary external harmonic potential required to assist the self-similar dynamics, changing character from expulsive to attractive along the expansion process. The opposite sequence is required for a compression. The transient expulsive potential is responsible for the speed-up of the process.  
\label{traj}  
}
\end{figure}
The set of boundary conditions at $t=0,\tau$ excludes the possibility of a linear ramp, as well as the family of trajectories, $\xi(t)=[at^2+bt+c]^{\frac{1}{2}}$, considered so far in the literature \cite{boxlaws1,boxlaws2,boxlaws3,bookinv}. However, it suffices to determine a polynomial ansatz for the trajectory $\xi(t)=\sum_{\ell} a_{\ell}t^{\ell}$, i. e. a scaling factor of the form
$\gamma(t)=\frac{\xi(t)}{\xi_0}=1 + [\gamma(\tau)-1] s^3 [10 + 3 s (2 s-5)]$,
 with $s=t/\tau$. This further determines the required time-dependent frequency of the auxiliary harmonic potential $U^{\rm aux}(x,t)$ according to Eq. (\ref{Uaux}),
\beqa
\Omega^2(t)=-\frac{\ddot{\xi}(t)}{\xi(t)}=-\frac{\gamma(\tau)-1}{\tau^2\gamma(t)} 60s [1 + s (2 s-3)].
\eeqa
The trajectory, displayed in Fig. 1 shows that during an expansion $U^{\rm aux}(x,t)$ becomes an expulsive potential in an early stage ($t<\tau/2$), providing the speed-up required to achieve the STA in an arbitrary finite time $\tau$ (bounds in the presence of perturbations will be discussed below). In a subsequent stage, $t>\tau/2$,  $\Omega^2(t)$ changes sign and $U^{\rm aux}(x,t)$ becomes a trapping potential, slowing down the expanding mode and reducing it to an eigenstate of the final Hamiltonian at $t=\tau$.
Precisely the opposite behavior is exhibited during a fast non-adiabatic compression.
Provided that an arbitrary $\Omega^2(t)$ dependence can be implemented, a STA has no lower bound for $\tau$ (notice however that $\Om(t)\sim\tau^{-1}$).
By contrast, the adiabatic condition 
\beqa
\max_t\max_{n,k}\bigg|\frac{\hbar\la\phi_n(t)|\partial_t|\phi_k(t)\ra}{(E_n(t)-E_k(t))}\bigg|\ll 1, 
\eeqa
leads to the requirement $mnk\xi(t)\dot{\xi}(t)/\hbar\ll 1$.

Note that the energy of the expanding mode
\beqa
\la \hat{H}(t)\ra_n=E_n(t)+\frac{m(\dot{\xi}^2-\xi\ddot{\xi})}{12}\left(2-\frac{3}{n^2\pi^2}\right),
\eeqa
has two contributions, the first one being the adiabatic energy $E_n(t)=E_n(0)\xi(0)^2/\xi(t)^2$, and the second one depending explicitly on $\{\dot{\xi},\ddot{\xi}\}$, so that $\la \hat{H}(\tau)\ra_n=E_n(\tau)$ given that a STA demands $\dot{\xi}(\tau)=\ddot{\xi}(\tau)=0$. This relation illustrates the fact that a STA is associated with a non-adiabatic evolution, which reproduces the adiabatic result at the end of the process.
Moreover, STA work as well for excited states: the time evolution of the $n$-th eigenstate of the initial trap leads to the $n$-th eigenstate of the final trap at $t=\tau$.    
As a result, STA in boxes pave the way for fast population-preserving cooling in the following sense.
Given a system described by the canonical ensemble with a density matrix $e^{-\beta\hat{H}}/\tr[e^{-\beta\hat{H}}]$, where $\beta=1/k_BT$, $k_B$ the Boltzmann constant and $T$ the temperature, the final temperature reads
\beqa
T(\tau)=\frac{T(0)}{\gamma(\tau)^2}.
\eeqa

%
%
\begin{figure}[t]
\begin{center}
\includegraphics[width=0.95\linewidth]{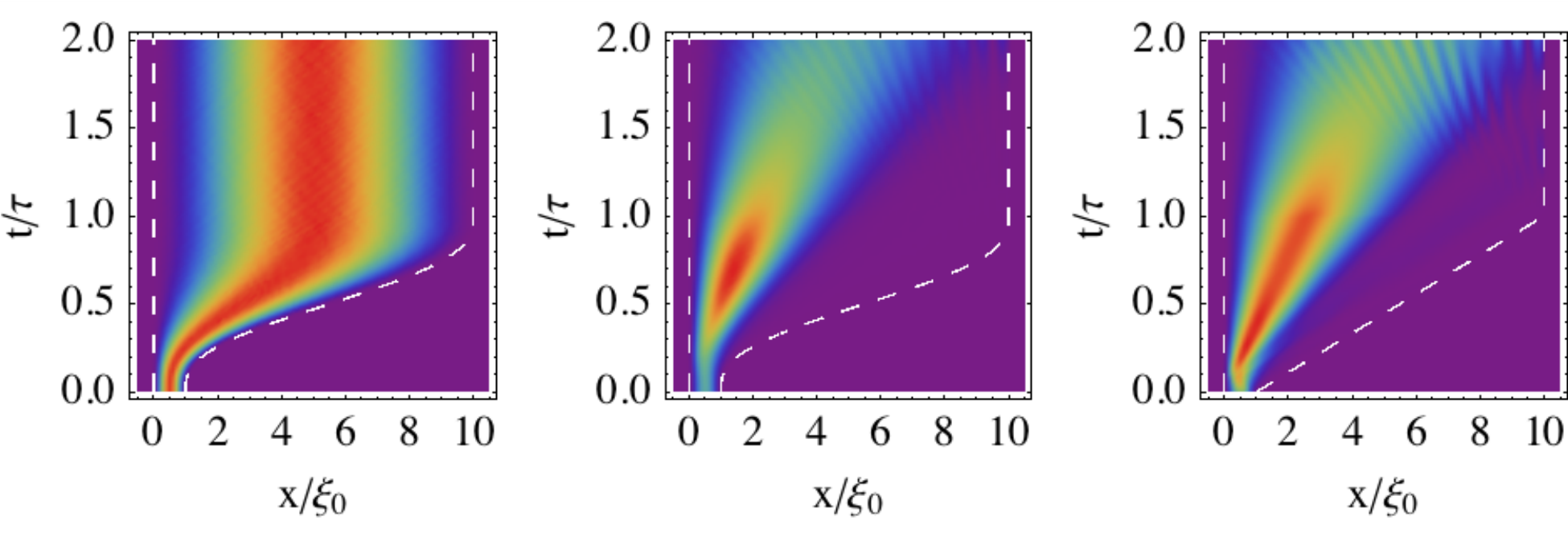}
\end{center}
\caption{\label{spsa} Space-time density plot of the scaled density profile $n(x,t)\gamma(t)$ during a shortcut to an adiabatic expansion 
for the ground-state mode of a box with smooth boundaries. A high fidelity-state results from the STA (left) in a box created with Gaussian barriers of height $V_0=10^4\hbar^2/m\xi_0^2$ and $\sigma=\xi_0/20$ ($\gamma=10$). In the absence of $U^{\rm aux}(x)$ (center), and for a linear ramp box width $\xi(t)$ (right) the cloud is excited and exhibits fringes associated with the reflections of the boundary. The color coding varies from purple to red as the function value increases. The dashed line follows the trajectory of the moving wall $\xi(t)$.
}
\end{figure}
%
These results strictly hold for a box with infinite walls at $x\in[0,\xi(t)]$. 
However, in the all-optical trap reported in \cite{becbox}, the end-cap lasers providing the box walls 
have a Gaussian profile which smooths the potential in a length scale $\sigma$,  i.e. a box trap of the form 
\beqa
U^{\rm box}(x)\!=\!V_0\big[\exp\!\left(\!-\frac{x^2}{2\sigma^2}\right)\!+\! \exp\!\left(\!-\frac{(x-\xi(t))^2}{2\sigma^2}\right)\big].
\eeqa 
A similar smoothing is expected in other physical realizations \cite{painters}.  Since this smoothing is expected to be the most significant deviation of laboratory potentials from the ideal infinite box, it is important to consider its effect on the self-similar dynamics required for a STA.  This can be quantified by the overlap  between the states resulting from the expansion in an idealized  and realistic box trap,  $|\Psi_n(t)\ra$ and $|\Psi_{n\sigma}(t)\ra$ respectively.
Clearly the role of $\sigma$ decreases (increases) during a expansion (contraction) of the box.
The numerical solution of the time-dependent Schr\"odinger equation for this box potential  is shown in Fig. 2, where the STA is compared with both the polynomial and linear expansion of the box in the absence of $U^{\rm aux}(x,t)$. It is seen that the STA is the only successful strategy and that the process is robust even for a substantial smoothing of the potential barriers, where the target states deviate from those of an idealized box. 

We have further explored numerically the dynamics under  ``concatenated STA'', in which the overall expansion is splitted into a sequence of $k$ STA with either constant expansion factor $\gamma$
or constant box size increment between consecutive steps. The efficiency of the process exhibits a non-monotonic improvement with increaisng $k$, suggesting a natural scenario where to combine STA techniques with optimal quantum control \cite{control1,control2}.


\noindent
{\bf Beyond adiabatic invariants: Shortcuts to adiabaticity in interacting many-body systems}

Knowledge of the adiabatic invariants for a single-particle in a time-dependent box has provided us with the insight 
of assisting the dynamics with an auxiliary potential
to design STA for expansions and compressions in a finite time, without inducing quantum transients associated with diffraction in time, which are ubiquitous in this type of scenario \cite{DGCM09}.
The technique, can be directly applied to non-interacting gases and other many-body quantum fluids which can be mapped to non-interacting systems. It further suppresses the Talbot dynamics associated with quantum carpets woven by the density profile typically observed in boxes \cite{Schleich1},
and the question naturally arises as to its applicability to interacting systems \cite{Schleich2}.
The presence of interactions, e.g. a two-body potential,
hinders the exploitation of the superposition principle in terms of the eigenmodes of the Lewis-Riesenfeld invariant. However, we note that to design a STA it suffices to enforce a self-similar dynamics and ultimately, no knowledge of adiabatic invariants is required.
As a result, we next consider a broader family of many-body systems, confined in a box, defined by the Hamiltonian
\beqa
\label{mbh}
\hat{\mathcal{H}}=\!
\sum_{i=1}^N\!\Big[\!-\frac{\hbar^2}{2m}\Delta_{\q_i}
\!+\!U^{\rm aux}(\q_i,t)\Big]\!+\!\epsilon\!\sum_{i<j}V(\q_i-\q_j)
\eeqa
where ${{\bf q}_i}\in\mathbb{R}^D$, $\Delta_{\q_i}$ is the Laplace operator in dimension $D$, the auxiliary term is now given by 
\beqa
U^{\rm aux}(\q,t)=-\frac{1}{2}m\frac{\ddot{\xi}(t)}{\xi(t)}|\q|^2,
\eeqa
 and the two-body interaction potential obeys ${\rm V}(\lambda \q)=\lambda^{-\alpha}{\rm V}(\q)$, e.g. for the Fermi-Huang pseudopotential describing s-wave scattering in ultracold gases, $\alpha=D$. For a hard-wall box, $r_i=|\q|_i\in[0,\xi(t)]$; we shall relax below this approximation and consider realistic potential boxes as those created in all-optical setups .
The case $D=1$ corresponds to a box with one-wall moving (the symmetric case in which both walls move in opposite directions can be obtained by a Duru transformation \cite{Duru89}).
For $D=2,3$ cylindrical and spherical symmetry is assumed respectively.
Without loss of generality, we choose the dimensionless time-dependent coupling constant $\epsilon=\epsilon(t)$ to satisfy $\epsilon(0)=1$. A stationary state  $\Psi(t)=\Psi(\q_1,\dots,\q_N;t)$ of $N$ particles and chemical potential $\mu$ follows for $t>0$ the evolution
\beqa
\label{mbphit}
\Psi(t)&=&\gamma^{-\frac{ND}{2}}
\exp\bigg[i\sum_{j=1}^N\!\!\frac{m|\q|_j^2\dot{\gamma}}{2\gamma\hbar}-i\frac{\mu \eta(t)}{\hbar}\bigg]\nonumber\\
& & \times \Psi\Big[\frac{\q_1}{\gamma(t)},\dots,\frac{\q_N}{\gamma(t)};0\Big],
\eeqa
with the boundary conditions $\Psi(t)=0$ for $|\q|_i=\xi(t)$ ($i=1,\dots,N$, in addition to $\Psi(t)=0$ for  $|\q|_i=0$ in $D=1$), as long as 
\beqa
\epsilon(t)=\gamma(t)^{\alpha -2},
\eeqa
which can be implemented exploiting a Feshbach resonance or modulating the transverse confinement in anisotropic systems \cite{Staliunas04,delcampo10}.

The self-similar dynamics in a STA leads to a scaling of all local correlation functions.
In particular the density of a given many-body state follows the law
$n(\q,t)=\int d\q_2\dots d\q_N|\Psi(\q,\q_2,\dots,\q_N;t)|^2=n[\q/\gamma(t),0]/\gamma(t)^D$.  By contrast, non-local correlation functions exhibit a non-trivial dynamics.
The one-body reduced density matrix $\varrho_1(\q,\q';t)=N\int d\q_2\dots d\q_N\Psi(\q,\q_2\dots,\q_N;t)\Psi^*(\q',\q_2\dots,\q_N;t)$ of a state obeying Eq. (\ref{mbphit}), follows the scaling law $\varrho_1(\q,\q';t)=\exp\big[i\frac{m(|\q|^2-|\q'|^2)\dot{\gamma}}{2\gamma\hbar}\big]\varrho_1(\q/\gamma,\q'/\gamma;0)$, analogous to that observed under harmonic confinement \cite{Demler}. The additional phase factor induces a major distortion of the momentum distribution, $n(\k,t)=\int d\q d\q'\varrho_1(\q,\q';t) \exp[i\k\cdot(\q-\q')]/(2\pi)^D$. However, a STA ensures that at $t=\tau$ the Lewis-Riesenfeld phase factor vanishes, so that the final state exhibits the same correlations of the initial state scaled by a factor $\gamma(\tau)$, 
\beqa
\varrho_1(\q,\q';\tau)&=&\frac{1}{\gamma(\tau)^D}\varrho_1[\frac{\q}{\gamma(\tau)},\frac{\q'}{\gamma(\tau)};0], \nonumber\\
n(\k,\tau)&=&\gamma(\tau)^Dn[\gamma(\tau) \k,0].
\eeqa

\noindent
{\bf Examples}

In the following we shall illustrate different aspects of shortcuts to adiabaticity in some paradigmatic models.


\begin{figure}[t]
\includegraphics[width=0.85\linewidth]{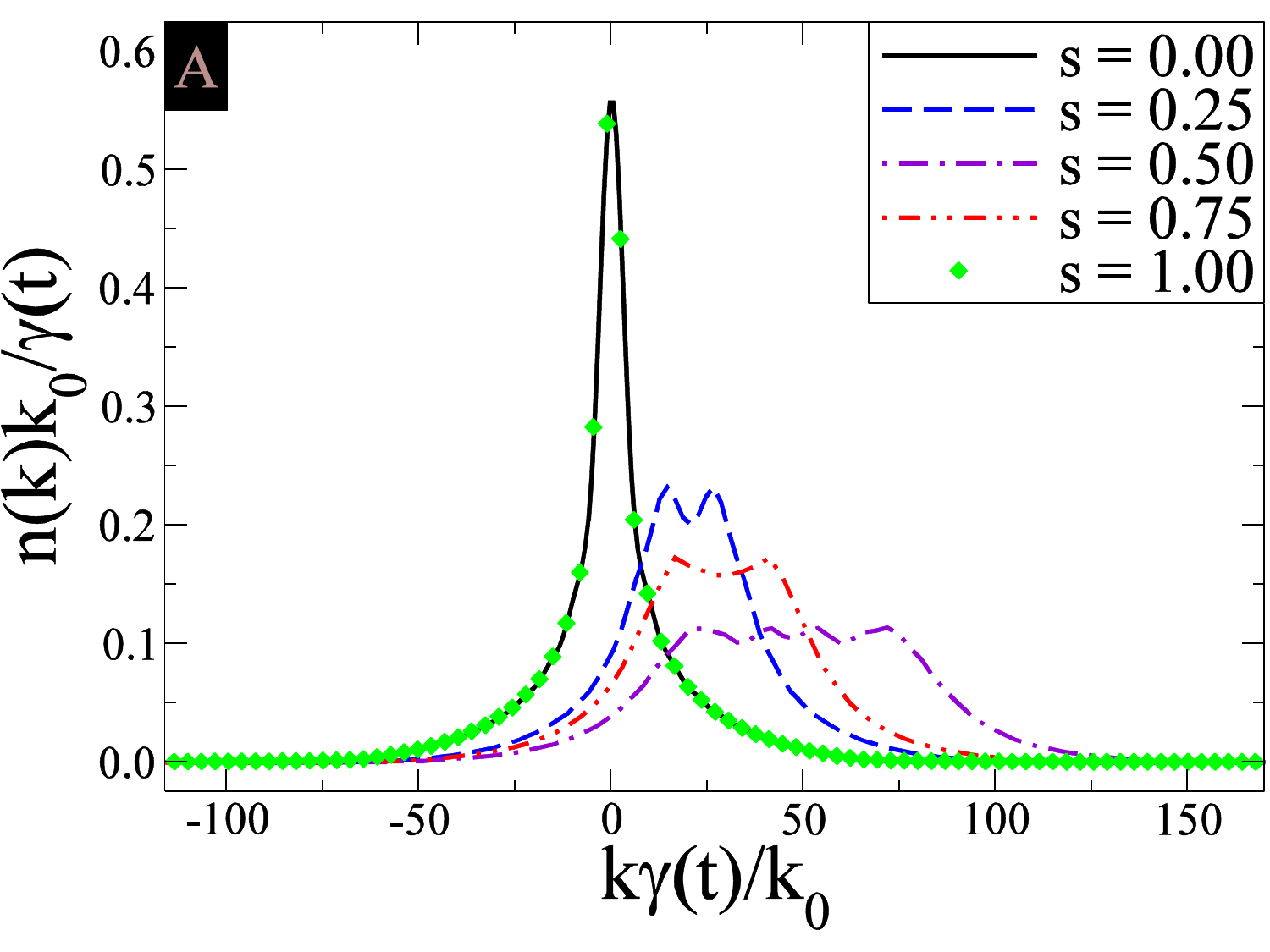}
\includegraphics[width=0.85\linewidth]{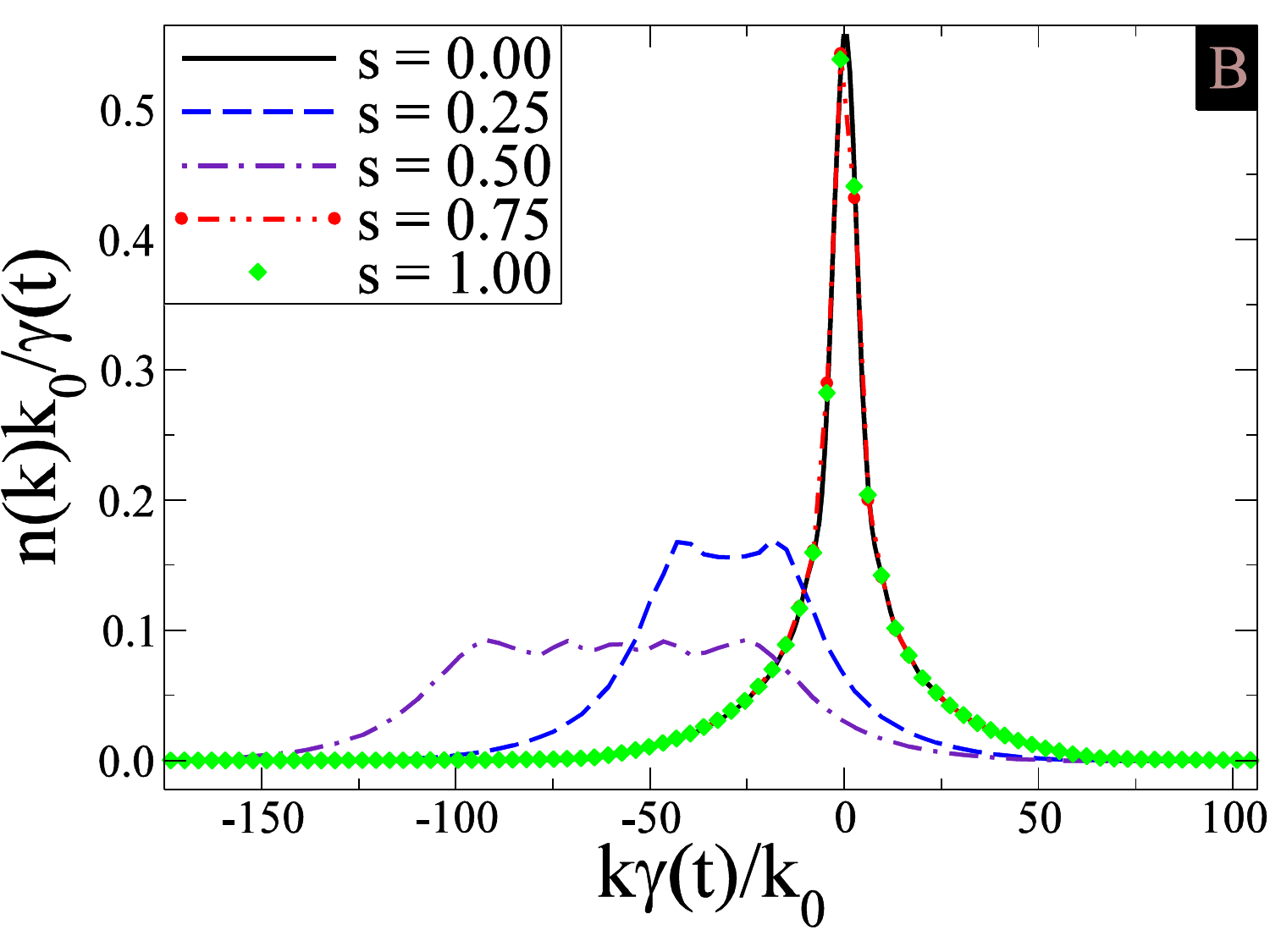}
\caption{Time evolution of the momentum distribution of a Tonks-Girardeau gas confined in box during an ultrafast expansion and compression providing a shortcut to adiabaticity. 
In a shortcut to an adiabatic expansion (A), due to the auxiliary external potential in Eq. (\ref{Uaux}), the cloud is accelerated in the first half of the process. At $s=1/2$ there is a partial mapping of the momentum distribution to the density profile of the initial state of the cloud. In the second half of the process, the cloud is slowed down and at $t=\tau$ it matches the result from an adiabatic dynamics: the state exhibits exactly the same correlations of the initial state, scaled up by the expansion factor $\gamma(\tau)=10$ ($N=10$). The reverse sequence is observed in a shortcut to an adiabatic compression (B),  where the target state is reached with high-fidelity already for $s=t/\tau=0.75$, after which the dynamics is actually adiabatic ($\gamma(\tau)=1/10$).
\label{satg}
}
\end{figure}
We shall first consider the evolution of correlations in a one-dimensional cloud of ultracold bosons in the limit of hard-core contact interactions, this is, in the Tonks-Girardeau (TG) regime \cite{Girardeau60}.
This system, as well as its lattice-version, has become a favorite test-bed to study the breakdown of thermalization and adiabaticity \cite{PSSV11}. Its many-body ground state is given  by the Bose-Fermi mapping \cite{Girardeau60}, $\Psi_{\rm TG}(q_{1},\dots,q_{N}) =\frac{1}{\sqrt{N!}}\prod_{1\leq j<k\leq N}\epsilon(q_{k}-q_{j})\det_{j,k=1}^{N}[\Psi_{j}(q_{k})]$, where $\epsilon(q)=1$ $(-1)$ if $x>0$ $(<0)$ and $\epsilon(0)=0$. In a STA, the self-similar dynamics is inherited from the single-particle orbitals $\Psi_{j}(q_{k},t)$ whence it follows that no tuning of interactions is required. 
Its density follows a scaling law for all $t$. The same holds true for the entanglement entropy with respect to a bipartition $[0,a\xi(t)]$ ($a<1$) \cite{CMV11}. The self-similar dynamics breaks down for the momentum distribution, which can be computed efficiently following \cite{Hrvoje}, and  we shall focus on its evolution along a STA. Different snapshots are depicted in Fig. 3A, and confirm that during an expansion the cloud is accelerated during the interval $[0,\tau/2]$ and slowed down during $[\tau/2,\tau]$. The reverse sequence, is observed in a fast frictionless compression. The axis are scaled up by the expansion factor $\gamma(t)$ in such a way that for an adiabatic dynamics, curves at different times would collapse into a single curve. 
Along a STA, the width and mean of the momentum distribution do not remain constant and change along the process.


A similar distortion of correlations, known as dynamical fermionization,  occurs in the dynamics of a cloud suddenly released from an arbitrary trap  \cite{1dbgases}. Under ballistic dynamics the asymptotic momentum distribution in a 1D expansion evolves to that of the dual system, a spin polarized Fermi gas. In particular for a cloud released from a box the exact time-evolution is not self-similar \cite{DM06} but dynamical fermionization is observed \cite{delcampo08,Buljan2}.
However, under a self-similar scaling law, the asymptotic $n(k)$ maps to the density profile of the initial state  \cite{Demler} and no dynamical fermionization occurs. 
 This is the case of relevance to STA, where the dynamical scaling law in Eq. (\ref{mbphit}) holds. (We note that the case of the initial harmonic confinement is singular in that the free expansion is self-similar and that  the the single-particle eigenstates can be written in terms of Hermite polynomials, which are eigenfunctions of the continuous Fourier transform. As a result the asymptotic momentum distribution can be related to both the initial density profile and the momentum distrubtion of non-interacting fermions. See \cite{delcampo11} for a discussion of STA in harmonic traps.) 
Moreover, this distortion of correlations is not restricted to expansion processes.  Along a STA, this is shown in Fig. 3 for both  expansions (A) and compressions (B). This is a spurious effect for the purpose of STA, which is to reproduce the adiabatic result in a finite short time. Indeed, the distortion  induced during the first half of the STA associated with the accelerated expansion or compression, is compensated in the second half of the dynamics, in such a way that the correlations of the initial state are reconstructed at $t=\tau$ and scaled by a factor $\gamma(\tau)$.

We next turn our attention to the design of STA  for  a BEC  in time-dependent box trap, where different strategies can be adopted depending on the dimensionality and the regime of interactions.
The time-dependent Gross-Pitaevski equation (TDGPE) governs the evolution of the normalized condensate wavefunction $\Phi(\q,t)$,
\beqa
i\hbar\partial_t\Phi(\q,t)\!=\!\big[\!-\frac{\hbar^2}{2m}\Delta_{\q}\!+\!U^{\rm aux}(\q,t)
\!+\!g_{D}|\Phi(\q,t)|^2\big]\Phi(\q,t),\nonumber\\
\eeqa 
with $|\q|\in[0,\xi(t)]$, for which adiabaticity conditions have been reported in \cite{BMT02}.
The ansatz 
\beqa
\Phi(t)=\gamma^{-\frac{D}{2}}\exp\bigg[i\frac{m|\q|^2\dot{\gamma}}{2\gamma\hbar}-i\frac{\mu \eta(t)}{\hbar}\bigg]\Phi\big[\q/\gamma(t);0\big],\nonumber\\
\eeqa
satisfies the TDGPE provided that 
\beqa
\eta(t)=\int^t\frac{dt'}{\gamma(t')^2},\quad \Om^2(t)=-\frac{\ddot{\xi}}{\xi},\quad g_D(t)=\frac{g_D(0)}{\gamma^{2-D}}.\nonumber\\
\label{ceq}
\eeqa

%
\begin{figure}[t]
\begin{center}
\includegraphics[width=0.95\linewidth]{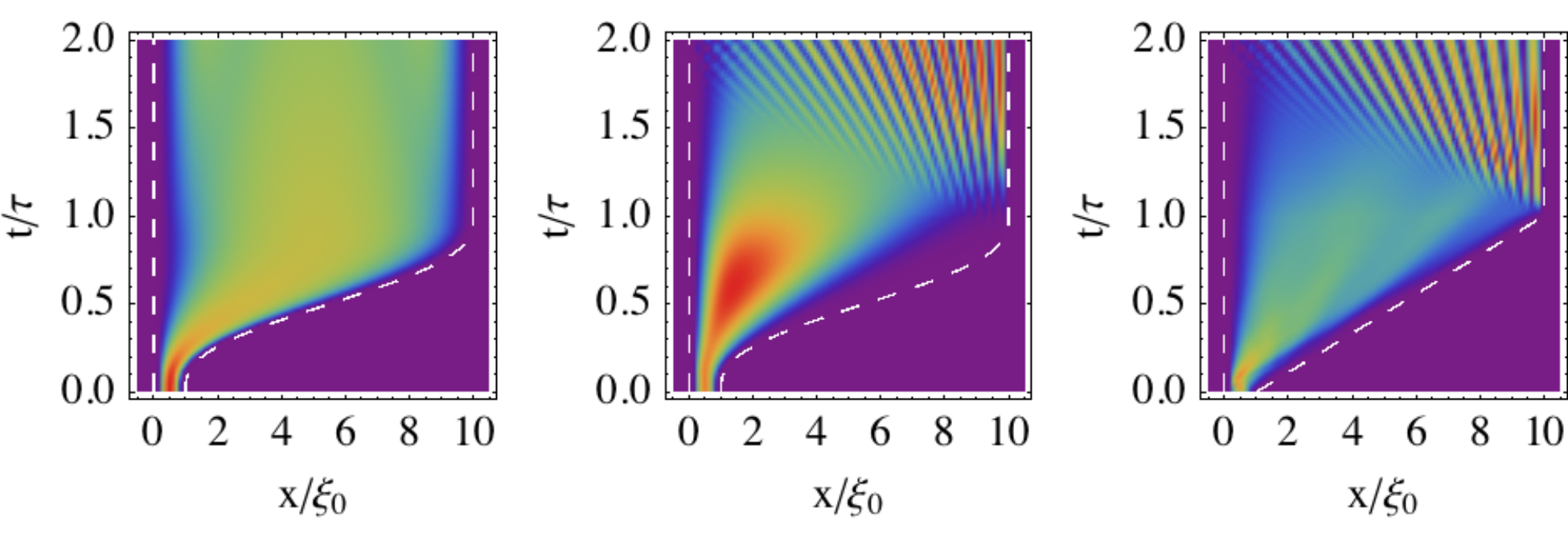}
\includegraphics[width=0.95\linewidth]{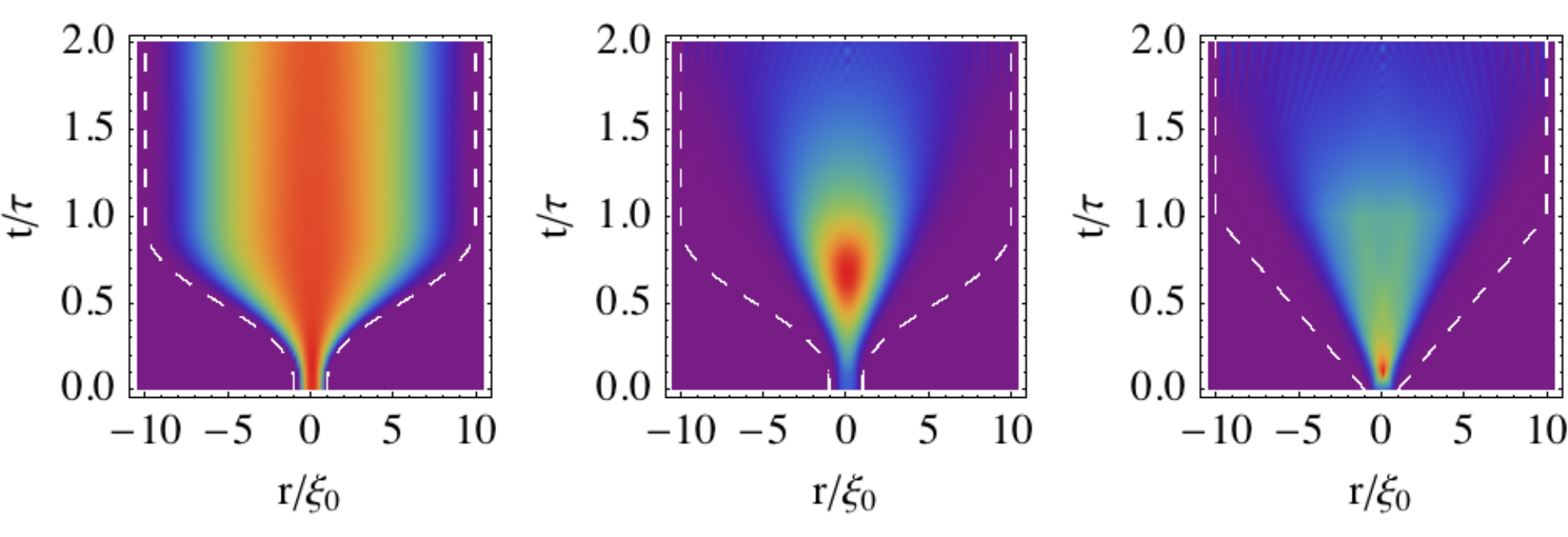}
\end{center}
\caption{Shortcut to an adiabatic expansion of a Bose-Einstein condensate.
The upper row shows  the space-time evolution of a quasi-1D BEC ($n(x,t)\gamma(t)$) while the lower row  corresponds to a quasi-2D ($n(r,t)\gamma(t)^2$) BEC cloud. 
The shortcut to adiabaticity (left) is compared with a polynomial (center) and a linear ramp (right) of the box width $\xi(t)$ in the absence of the auxiliary harmonic potential 
 ($\sigma/\xi_0=1/20$, $g_{D}=10\hbar^2/m\xi_0^{2-D}$,  and  $\gamma=10$).
\label{figsabec}}
\end{figure}
These relations constitute the box analogue of the well-known Castin-Dum-Kagan-Surkov-Shlyapnikov relations in harmonic traps \cite{CD96,KSS96}.
The two-dimensional case is special since the scaling law holds when $g_{2D}(t)$ is kept constant.

Figure 4 is a set of numerical solutions of the time-dependent Gross-Pitaeveskii equation that illustrate the robustness of the STA for realistic BEC experiments.
We consider a box trap with Gaussian barriers and in all numerical simulations interactions are kept constant, i.e.  $g_D(t)=g_D(0)$, deviating from 
the ideal  prescription in Eq. (\ref{ceq}). The top row illustrates the dynamics for a quasi-1D BEC. The (one-body) fidelity between the resulting state $\Phi(\tau)$ 
and the ground state of the final box $\Phi_{gs}^{f}$ is $\mathcal{F}_{\gamma=10}=|\la\Phi(\tau)|\Phi_{gs}^f\ra|^2=0.911$. For smaller values of $\gamma$,  the fidelity is even higher $\mathcal{F}_{\gamma=3}=0.999$ as expected, given that implementation of the exact STA requires a smaller tuning of $g_1$. The bottom row shows the dynamics of a quasi-2D cloud, which requires no interaction tuning in a STA, but is more sensitive to the smoothness of the box boundaries, $\mathcal{F}_{\gamma=10}=0.988$. 

It is noteworthy that in the Thomas-Fermi regime, the kinetic term contribution can be neglected and it is possible to induce an exact self-similar dynamics (and a STA)  
exclusively with the help of an external field. Then, the scaling ansatz is a solution of the TDGPE provided that
\beqa
\eta(t)=\int^t\frac{dt'}{\gamma(t')^D},\quad \Om^2(t)=-\frac{\ddot{\xi}}{\xi},\quad g_D(t)=g_D(0).\nonumber\\
\eeqa
This regime is particularly robust against the smooth boundaries of physically realizable box potentials.
The simulations correspond to the most delicate regime with moderate mean-field interactions, both far from the non-interacting and Thomas-Fermi limits.

\noindent
{\bf Discussion}

In conclusion, we have presented a method to drive an ultrafast dynamics in a time-dependent box trap which
reproduces the adiabatic result at the end of the evolution.
The method is assisted by an auxiliary external harmonic
potential which provides the speed-up and is applicable to a large family of both non-interacting and  interacting many-body systems supporting dynamical scaling laws,
where it not only leads to a robust expansion of the density but also preserves the non-local correlation functions of the initial state,
up to an expansion factor. 
The  proposal is applicable to realistic box potentials and can be implemented in the laboratory with well-established technology.
Its applications range over all scenarios requiring a shortcut to adiabaticity, i.e., probing strongly correlated phases, preventing decoherence, the effect of perturbations and atomic losses.  
The method can be directly applied as well to ultrafast population-preserving cooling methods,  quantum heat engines and refrigerators \cite{Koslof} providing an alternative to the paradigmatic model of a quantum piston \cite{QJ12}.

\noindent
{\bf Acknowledgements:}\\
\noindent
It is a pleasure to thank H. Buljan, X. Chen, B. Damski, M. G. Raizen, E. Timmermans and W. H. Zurek for discussions and useful comments on the manuscript.
This work was supported by the U.S. Department of Energy through the LANL/LDRD
Program and a  LANL J. Robert Oppenheimer fellowship.

\vspace{0.21cm}

\end{document}